\documentclass[prc,twocolumn,showpacs,showkeys,unsortedaddress,nofootinbib]{revtex4}
\usepackage{graphicx}
\def\be{\begin{eqnarray}}
\def\ee{\end{eqnarray}}
\def\bc{\begin{center}}
\def\ec{\end{center}}

\def\rmF{{\rm F}}
\def\rmd{{\rm d}}

\def\om{\omega}

\newcommand{\lsim}{\stackrel{\scriptstyle <}{\phantom{}_{\sim}}}
\newcommand{\gsim}{\stackrel{\scriptstyle >}{\phantom{}_{\sim}}}
\begin{document}
\title{Neutrino emission due to Cooper-pair recombination in neutron stars revisited}
\author{E.E. Kolomeitsev}
\affiliation{GSI, Plankstra\ss{}e 1, D-64291 Darmstadt, Germany}
\thanks{present address}
\affiliation{School of Physics and Astronomy, University of
Minnesota, 116 Church Str. SE, Minneapolis 55455, USA}
\author{D.N. Voskresensky}
\affiliation{GSI, Plankstra\ss{}e 1, D-64291 Darmstadt, Germany}
\affiliation{Moscow Engineering Physical Institute,\\ Kashirskoe
  Avenue 31, RU-11549 Moscow, Russia}
\begin{abstract}
Neutrino emission in processes of breaking and formation of neutron and proton Cooper pairs is
calculated within the Larkin-Migdal-Leggett approach for a superfluid Fermi liquid. We demonstrate
explicitly that the Fermi-liquid renormalization respects the Ward identity and assures the weak
vector current conservation. The systematic expansion of the emissivities for small temperatures
and  nucleon Fermi velocity, $v_{\rmF,i}$, $i=n,p$ is performed. Both neutron and proton processes
are mainly controlled by the axial-vector current contributions, which are not strongly changed in
the superfluid matter. Thus, compared to earlier calculations the total emissivity of processes on
neutrons paired in the 1S$_0$ state is suppressed by a factor $\simeq (0.9\mbox{---}1.2)\,
v_{\rmF,n}^2$. A similar suppression factor ($\sim v_{\rmF,p}^2$) arises for processes on protons.
\end{abstract}
\date{\today}
\pacs{
13.15.+g,  
21.65.Cd,  
26.30.Jk,   
26.60.-c,   
71.10.Ay  
}
\keywords{neutron star, Fermi liquid, superfluidity, Cooper pair breaking, neutrino
emission,vertex function correction}

\maketitle

\section{Introduction}

In minutes/hours after the birth a neutron star   cools down to a temperature $T\sim $ MeV via a
neutrino transport to the surface and then becomes transparent for neutrinos. Hereafter during
$\sim 10^{5}$~yr the cooling is determined by emissivity of neutrinos produced in direct reactions
\cite{ST83,MSTV90,YLS99,V01,PGW,Sedr07}. For such temperatures neutrons and protons in the neutron
star interior are highly degenerate. Therefore the rate of the neutrino production is suppressed by
the  reaction phase space,  the stronger, the more nucleons are involved.

The most efficient are one-nucleon processes, e.g., $n\to pe\bar{\nu}$, called direct Urca (DU)
reactions. Their emissivity is $\varepsilon^{\rm DU}\sim 10^{27}\times T_9^6\,(n/n_0)^{2/3}\theta
(n -n_{c}^{\rm DU}) \,\frac{\rm erg}{\rm cm^3\, s} $, see ~\cite{LPPH91}, where $T_9=T/(10^9\, {\rm
K})$\,, $n$ is the nucleon density measured in  units of the nuclear matter saturation density
$n_0$. The DU processes are operative only when the proton fraction exceeds a critical value of
11--14\%. Equations of state constructed from realistic nucleon-nucleon interactions, like
Urbana-Argonne one~\cite{APR98}, show that this condition is fulfilled only at very high densities.
This implies that the DU processes may occur only in most heavy neutron stars, e.g., with masses
$\sim 2\,M_{\odot}$ for the equation of state \cite{APR98}, where $M_\odot=2\times 10^{33}$~g is
the solar mass. At $n\sim n_0$ the proton fraction is typically about 3--5\%, cf. Fig.~2 in
Ref.~\cite{Army}.

In the absence of the DU processes, most efficient become two-nucleon reactions, e.g., $nn\to
npe\bar{\nu}$, called modified Urca processes (MU) with the emissivity $\varepsilon^{\rm MU}\sim
10^{21}\times T_9^8\,(n/n_0)^{2/3} \,\frac{\rm erg}{\rm cm^3\, s}$\,, cf.~\cite{MUrca}.  Note the
smaller numerical prefactor and the higher power of the temperature for the emissivity of the
two-nucleon processes compared to the DU emissivity. In-medium change of the nucleon-nucleon
interaction in the spin-isospin particle-hole channel due to pion softening may strongly
increase the two-nucleon reaction rates, which, nevertheless, in all relevant cases remain
significantly smaller than that for the DU~\cite{VS86,MSTV90}. The nucleon bremstrahlung reactions,
like $nn\to nn\nu\bar{\nu}$ (nB) and $pp\to pp\nu\bar{\nu}$ (pB), have an order of magnitude
smaller emissivity than MU.

At low temperatures the nucleon matter is expected to undergo a phase transition into a state with
paired nucleons~\cite{Migdal-pair}. The neutron superfluidity and/or proton superconductivity take
place below some critical temperatures $T_{{\rm c},n}$ and $T_{{\rm c},p}$, respectively, which
depend on the density. At densities $n<(2\mbox{--}4)\,n_0$ neutrons are paired in the 1S$_0$ state
and  in the 3P$_2$ state at higher densities. Protons are paired in 1S$_0$ state for densities
$n\lsim (2\mbox{--}4)n_0$. Paring gaps, $\Delta_i$, are typically $\sim (0.1\mbox{--}1)$~MeV and
depend crucially on details of the interaction in the particle-particle channel, see Fig.~5
in~\cite{BGV04}.

The gap in the energy spectrum significantly reduces the phase space of the nucleon processes
roughly by the factor $\exp(-\, \Delta /T)$\, for the one-nucleon DU process and $\exp(-\, 2\Delta
/T)$\ for two-nucleon processes. However, even with inclusion of the nucleon pairing effects the DU
rate is large enough that the occurrence of these processes would lead to an unacceptably fast
cooling of a neutron star in disagreement with modern observational soft $X$-ray
data~\cite{BGV04,GV05,Army}. This statement has been tested with gaps varying in a broad band
allowed by different microscopic calculations. Certainly, DU processes could be  less efficient if
one kept gaps finite also at  high densities. Microscopic calculations do not support this
possibility. Thus, according to the recent analysis the DU processes most probably will not occur
in typical neutron stars with masses in the range of $1.0\mbox{--} 1.5~M_{\odot}$, based on the
cooling and the population syntheses scenarios~\cite{BGV04,GV05,BP}.

The superfluidity allows for a new mechanism of neutrino production associated with Cooper pair
breaking or formation (PBF), e.g., the reaction   $n\to n \, \nu\, \bar{\nu}$ and $p\to p \, \nu\,
\bar{\nu}$, where one of the nucleons is paired. For 1S$_0$ neutron pairing the PBF emissivity was
evaluated first in~\cite{FRS76} in the Bogolyubov $\psi$-operator technique and then
in~\cite{VS87,SV87} within the Fermi-liquid approach.

The proton PBF emissivity was estimated in~\cite{VS87} with account for in-medium renormalization
of the nucleon weak-interaction vertex due to strong interactions. Mixing of electromagnetic and
weak interactions through the electron--electron-hole loop can additionally change the proton
vertex~\cite{VKK,L00}. There are also relativistic corrections  to the axial-vector coupling
vertices of the order $v_{\rmF,i}^2$, $v_{\rmF,i}$ are neutron and proton Fermi velocities, $i=n,
p$, ~\cite{KHY}. These three effects together resulted in a one or two orders of magnitude
enhancement of the proton PBF emissivity compared to that evaluated with the free vector-current
vertex.  Thus, one concludes that neutron PBF and proton PBF emissivities can be equally important
for neutron star cooling depending on the parameter choice and, especially, on relation between
gaps $\Delta_p$ and $\Delta_n$. The PBF emissivity for the $3P_2$ neutron pairing has been analyzed
in~\cite{YKL99}.

Following \cite{VS87,SV87,MSTV90} the emissivity of the neutron PBF and proton PBF processes is
estimated as $\varepsilon^{\rm iPFB}\sim 10^{28} \times ({\Delta_i}/{\rm MeV})^7\,
({T}/{\Delta_i})^{1/2} (n_i/n_0)^{1/3}\, e^{-2\Delta_i /T}\,\, \frac{\rm erg}{\rm cm^3\, s}$\, for
$T\ll \Delta_i$ and $i=n,p$. Having a large numerical pre-factor and very moderate temperature
dependence of the pre-exponent, these reactions significantly contribute to the neutron star
cooling provided gaps are not too small. These processes have been included in the cooling code
rather recently~\cite{SVSWW97}. From that time  the PBF reactions are the main part of any
cooling-scenario together with the MU processes~\cite{Minimal,BGV04,GV05,Yakov}. Uncertainties in
the pairing gaps are large. Therefore surface temperatures of neutron stars computed in different
approaches vary significantly.

Kundu and Reddy \cite{KR} and Leinson and Perez ~\cite{LP} made an important observation: all
previous calculations of the neutrino reactions in superfluid matter disrespect the Ward identity
and, as a consequence, the conservation of the electro-weak vector current.

The Ward or in general case Ward-Takahashi identities impose non-trivial relations between vertex
functions and Green functions, which synchronize any modification of the Green-function with a
corresponding change in the vertex function. Satisfying these relations one assures that the
symmetry properties of the initial theory are preserved in actual calculations. For instance we
start with the theory of weak interactions with a conserved vector current. The current would
remain trivially conserved in calculations with only bare vertices and bare Green functions. In
strongly interacting systems the Green functions change necessarily, but for quasiparticle Green
functions the current conservation is easily restored by a proper inclusion of short-range
correlations in the vertices, cf. development of the Fermi-liquid theory by
Migdal~\cite{M63,M67,BP}. Following these two simplest cases, Refs.~\cite{FRS76,YKL99,KHY} did not
incorporate any medium effects, whereas~\cite{VS87} used dressed quasiparticle Green functions
together with dressed normal vertices. In case of the superfluid system situation is more peculiar.
Since in the superfluid system the nucleon Green function notoriously differs from the free one,
the vector-current vertex must get corrections even if no other interaction between quasiparticles
is included. Additional anomalous vector-current vertices disregarded in previous calculations must
be properly accounted for. These corrections {\em{cancel exactly the vector-current contributions}}
to the neutrino emissivity for zero neutrino momenta, cf. \cite{LP}.

Assuming that the axial-vector current contributes only little to the PBF emissivity,
Ref.~\cite{LP} claimed that the PBF emissivity calculated in~\cite{FRS76,VS87,YKL99} is to be
suppressed by factor $\sim v_{\rmF,n}^4 /20\sim  10^{-3}$ for $n\sim n_0$ for neutrons and by $\sim
10^{-7}$ for protons in the case of 1S$_0$ pairing. Such a severe reduction of the neutron and
proton PBF emissivities could significantly affect previous results on the neutron star cooling
dynamics. Ref.~\cite{SMS} revises results~\cite{LP} applying expansion in $\vec{q}^{\,2}$ parameter
and putting $v_{\rmF,n}=0$. Ref. \cite{SMS} claims that suppression factor for the neutron PBF
emissivity is $\sim T/m^*$, where $m^*$ is the nucleon effective mass. This would reduce the
neutron PBF emissivity by the factor $\sim 5\times 10^{-3}$  for temperatures $T\sim 0.5\,
T_{c,n}$, cf. Fig.~5 in~\cite{SMS}.

Refs. \cite{LP,SMS} used the convenient Nambu-Gorkov matrix formalism developed to describe
metallic superconductors~\cite{Nambu,Schriffer}. The price paid for the convenience is that the
formalism does not distinguish interactions in the particle-particle and particle-hole channels.
Such an approach is, generally speaking, not applicable to the strongly interacting matter present
in neutron stars. In the nucleon matter at low temperatures the  $nn$ and $pp$ nucleon-nucleon
interactions in the particle-particle channel are attractive, whereas in  relevant particle-hole
channels they are repulsive~\cite{M67,VS87,FL}. The adequate formalism was developed by Larkin and
Migdal for Fermi liquids with pairing at $T=0$ in Ref.~\cite{LM63} and generalized then by Leggett
for $T\neq 0$ in Ref.~\cite{Leg65a}.

In the present paper using Larkin-Migdal-Leggett formalism we analytically calculate neutrino
emissivity from the superfluid neutron star matter with the 1S$_0$ neutron-neutron and
proton-proton pairing. Both normal and anomalous vertex corrections are included. We explicitly
demonstrate that the Fermi-liquid renormalization~\cite{LM63} respects the Ward identity and vector
current conservation. Our final estimations of neutron and proton PBF emissivities differ from
those in~\cite{LP,SMS}. We find that the main term in the emissivity $\sim v_{\rmF,i}^2$
follows from the axial-vector current, whereas the leading term in the emissivity from the vector
current appears only at the $v_{\rmF,i}^4$ order, as in \cite{LP}.

In next section we present the general expression for the emissivity of the neutron PBF processes
formulated in terms of the imaginary part of the current-current correlator for weak processes on
neutral currents. Then within the Fermi-liquid approach to a superfluid we introduce  Green
functions and the gap equation. In Sect. 3 we formulate and solve Larkin-Migdal equations for
vertices and apply them to calculate the imaginary part of the current-current correlator. In Sect.
4 we calculate the neutrino emissivity of the neutron PBF processes. Also in this section and in
Appendix a comparison of   our results with results of previous works is performed. In Sect. 5 we
consider emissivity of the proton PBF processes. Conclusions are formulated in Sect.~6.

\section{General expressions: emissivity, Green functions and pairing gaps}

\subsection{Neutrino emissivity}

The weak neutrino-neutron and neutrino-proton interactions on
neutral currents are described by the effective low-energy
Lagrangian
\be
\mathcal{L}=\frac{G}{2\,\sqrt
2}\,\sum_{i=n,p}(V_{i}^\mu-A_{i}^\mu)\,l_\mu\,, \label{weak-lag}
\ee where $l_\mu=\bar{\nu}\gamma_\mu\, (1-\gamma_5)\, \nu$ is the
lepton current and $V^\mu_{i}=g_V^{(i)}\,\bar{\Psi}_i\,
\gamma^\mu\, \Psi_i$ and $A^\mu_{i}=g_A^{(i)}\,\bar{\Psi}_i\,
\gamma^\mu \gamma_5\, \Psi_i$ stand for nucleon (neutron or
proton) vector and axial-vector currents with nucleon bi-spinors
$\Psi_i$. The coupling constants are  $g_V^{(n)}=g_V=-1$,
$g_V^{(p)}=c_V=1-4\mbox{sin}^2 \theta_{\rm W}\simeq 0.04$ and
$g_A^{(p)}=-g_A^{(n)}=g_A=1.26$. The Fermi constant is $G\approx
1.2\times 10^{-5}$~GeV$^{-2}$. For the non-relativistic nucleons
$V_i^\mu\approx\psi_i^\dag\big({p}')\,(1,\,(\vec{p\,}'+\vec{p\,})/2\,m\big)\,
\psi_i (p)$ and $A^\mu\approx
\psi_i^\dag(p')\,\big(\vec{\sigma}(\vec{p\,}'+\vec{p\,})/2\,m
,\,\vec{\sigma}\big)\, \psi_i(p)$\,, where
$\vec{\sigma}=(\sigma_1,\sigma_2,\sigma_3)$ are the Pauli matrices
acting on nucleon spinors $\psi_i$, and $\vec{p\,}'$ and
$\vec{p\,}$ are outgoing and incoming momenta, $m$ is the mass of
the free nucleon, cf. \cite{EW}.

Neutrino emissivity for one neutrino species can be calculated as
\be
\varepsilon_{\nu\bar\nu}=\frac{G^2}{8}\intop
\frac{d^3 {q}_1}{(2\pi)^3\,2\, \om_1}
\frac{d^3 {q}_2}{(2\pi)^3\,2\, \om_2}\,
\om\,f_{\rm B}(\om)\,2\,\Im
\sum\chi(q), \label{emissivity}
 \ee
where $q=(\om,\vec{q}\,)=q_1+q_2$,  $q_{1,2}=(\om_{1,2},\vec{q}_{1,2})$
are 4-momenta of outgoing neutrino and antineutrino, $f_{\rm B}
(\om)=1/(\exp(\om/T)-1)$ are Bose occupations, and $\Im\chi$ is
the imaginary part of the susceptibility of the nucleon matter to
weak interactions, i.e. the Fourier-transform of the
current-current correlator $\langle\big(V_\mu(x) l^\mu(x)-A_\mu(x)
l^\mu(x)\big) \big(V^\nu(y)
l_\nu^\dag(y)-A^\nu(y)l_\nu^\dag(y)\big)\rangle$, for weak
processes. The sum in (\ref{emissivity}) is taken over the lepton
spins.

According to the optical theorem  $\Im\chi$ can be expressed
as the sum squared matrix elements of all available reactions with
all possible intermediate states, $\sum |M|^2$. A particular
contribution to $\sum |M|^2$ can be also calculated within the
Bogolubov $\psi$-operator approach for a given form of the
nucleon-nucleon interaction, as it has been done in
Refs.~\cite{FRS76,YKL99}. In this approach, however, an account of
further in-medium modifications of nucleon propagators and
interaction vertices is obscured by a danger of double counting.
The Green-function technique for Fermi liquid~\cite{M67,FL} is
more suitable for such extensions as it was demonstrated
in~\cite{VS87,VS86}. We will follow the Green-function approach
for superfluid Fermi liquids~\cite{LM63,M67}. As a simplification
we will focus on the low temperature limit $T\ll \Delta$. The
temperature dependence enters through nucleon occupation factors
$\propto e^{-\Delta/T}(1+O(T^2/\Delta^2)$ and also as
$1+O(T^2/\epsilon_{\rm F}^2)$ corrections in the low-temperature
expansion of standard Fermi integrals, when the high energy region
$\epsilon \gg \Delta$ is dominating in the integrals. Since the
boson occupation factor $f_B$ in (\ref{emissivity}) generates
already the leading exponent $e^{-2\Delta/T}$ we can evaluate
$\Im\chi$ for $T=0$, see also discussion below.

\subsection{Nucleon Green functions and pairing gaps}

The nucleon Green function for the interacting system in a normal state ("n.s."), i.e.  without
paring, is given by the Schwinger-Dyson equation, which in the momentum representation reads
\be
\widehat{G}_{\rm n.s.}(p)=\widehat{G}_0(p) +\widehat{G}_0(p)\, \widehat{\Sigma}_{\rm n.s.}(p) \,
\widehat{G}_{\rm n.s.}(p)\,
\nonumber\ee
with
$\widehat{G}_0(p)=G_0(\epsilon,\vec{p})\,\hat{\mathbf{1}}=\hat{\mathbf{1}}/(\epsilon-\epsilon_p+i\,
0\,{\rm sng}\,\epsilon)$\,, where $\hat{\mathbf{1}}$ is the unity matrix in the spin space. All
information on the interaction is incorporated in the nucleon self-energy $\widehat{\Sigma}_{\rm
n.s.}$, being a functional of the Green functions $\widehat{G}_{\rm n.s.}$. In  absence of the
spin-orbit interaction the full Green function is also diagonal in the spin space, i.e.
$\widehat{G}_{\rm n.s.}=G_{\rm n.s.}\,\hat{\mathbf{1}}$\,. For strongly interacting systems, like
dense nucleon matter, the exact calculation of $G_{\rm n.s.}$ is extremely difficult task. However,
for strongly degenerate nucleon systems at temperatures, $T$, much less than the neutron and proton
Fermi energies, $\epsilon_{{\rm F},i}$, $i=n,p$, fermions are only slightly excited above the Fermi
see. So, the full Green function of  the normal-state is given by the sum of the pole term and a
regular part:
\be
G_{\rm n.s.}(p)=\frac{a}{\epsilon-\epsilon_p+i\,0\, {\rm
sgn}\epsilon} + G_{\rm reg}(p)\,,
\label{GF}
\ee
where the excitation energy is counted from the nucleon chemical potential $\mu$, $\epsilon_p
={p^2}/{(2\,m^*)}-\mu$\,, $\mu \simeq\epsilon_{{\rm F}}=p_{{\rm F}}^2/(2m^*)$ for low temperatures
under consideration, $p_{{\rm F}}$ is the Fermi momentum. The effective mass and the non-trivial
pole residue are determined by the real part of the self-energy, as $a^{-1}=1-\left({\partial
\Re\Sigma_{\rm n.s.}}/{\partial \epsilon}\right)_\rmF$ and $1/m^* =a\, \left({1}/{m} + 2\,
{\partial \Re\Sigma_{\rm n.s.}}/{\partial p^2} \right)_\rmF$. The subscript $"{\scriptsize \rmF}"$
indicates that the corresponding quantities are evaluated at the Fermi surface ($\epsilon ,
\epsilon_p\to 0$). According to Ref.~\cite{M67} only the pole part of $G_{\rm n.s.}$ is relevant
for the description of processes happening in a weakly excited Fermi system. The regular part can
be absorbed by the renormalization of the particle-particle and particle-hole interactions at the
Fermi surface. The quantities $m^*$ and $a$ can be expressed through the Landau-Migdal parameters
characterizing the fermion interaction at the Fermi surface at zero energy-momentum transfer. The
imaginary part of the self-energy, $\Im\Sigma_{\rm n.s.}$, can be omitted in the pole term of the
Green function (\ref{GF}) in the low-temperature limit (quasiparticle approximation).

In a system with pairing a new kind of processes such as transition of a particle into a hole and a
condensate pair and vice versa become possible. The one-particle one-hole irreducible amplitudes of
such processes can be depicted as \cite{LM63}
\be
-i\,
\hat{\Delta}^{(1)}={\includegraphics[width=1.5cm]{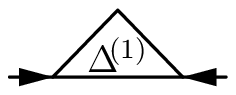}}
\,,\quad -i\,
\hat{\Delta}^{(2)}={\includegraphics[width=1.5cm]{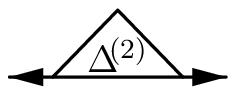}}\,.
\label{delta} \ee Besides the "normal" Green functions shown by
the "thick" line
$i\,\hat{G}=\parbox{1cm}{\includegraphics[width=1cm]{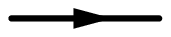}}$\,
one introduces "anomalous" Green functions
$
i\hat{F}^{(1)}=\parbox{1cm}{\includegraphics[width=1cm]{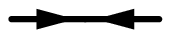}}$\,,
and
$i\hat{F}^{(2)}=\parbox{1cm}{\includegraphics[width=1cm]{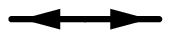}}$\,.
The full normal and anomalous Green functions are related by
Gor'kov equations
\be
&&\hat{G}(p)=\hat{G}_{\rm n.s.}(p)+\hat{G}_{\rm n.s.}(p)\,\hat{\Delta}^{(1)}(p)\, \hat{F}^{(2)}(p)\,,
\nonumber\\
&&\hat{F}^{(2)}(p)=\hat{G}^{h}_{\rm n.s.} (p)\, \hat{\Delta}^{(2)}(p)\,\hat{G}(p)\,.
\label{Gorkov}
 \ee
The second equation involves the normal-state Green function of
the hole (superscript $"h"$), which in the absence of a spin-orbit
interaction is simply $i\,\hat{G}^{h}_{\rm n.s.}
(p)=i\,\hat{G}_{\rm n.s.} (-p)=
\parbox{1cm}{\includegraphics[width=1cm,angle=180]{Gn.eps}}$\,.
In the case of the $1S_0$ pairing the spin structures of the anomalous Green functions and the
transition amplitudes (\ref{delta}) are simple: $\hat{\Delta}^{(1)}=\hat{\Delta}^{(2)}=\Delta\, i\,
\sigma_2$ and $\hat{F}^{(1)}=\hat{F}^{(2)}=F\, i\sigma_2$\,. Eqs.~(\ref{Gorkov}) are to be
completed by the equation for the amplitude $\hat{\Delta}^{(1)}(p)$,
\be
\big[\hat{\Delta}^{(1)}\big]^a_b=\!\!\intop\!\!\frac{d^4 p'}{(2\pi)^4
i}\big[\widehat{V}(p,p')\big]^{ac}_{bd}
\big[\hat{G}(p')\hat{\Delta}^{(1)}\!(p')\hat{G}_{\rm n.s.}^h(p')
\big]^d_c ,\label{gapeq1}
\ee
where $\widehat{V}$ stands for a two-particle irreducible potential, which determines the full
in-medium particle-particle scattering amplitude. The potential $\widehat{V}$ can be separated in
the scalar and spin-spin interactions  defined as
\be
\big[\widehat{V}\big]^{ac}_{bd}=V_V\,(i\sigma_2)^a_b\, (i\sigma_2)^c_d
+ V_A\, (i\sigma_2\vec{\sigma})^a_b\, (\vec{\sigma}\, i\sigma_2)^c_d\,.
\nonumber\ee
Solution of the Gor'kov equations (\ref{Gorkov}) is straightforward. The relevant pole parts of
Green functions are
\be
G(p)= \frac{a\,(\epsilon+\epsilon_p)}{\epsilon^2-E_p^2+i0{\rm
sgn}\epsilon}, F(p)=\frac{-a\, \Delta}{\epsilon^2-E_p^2+i0{\rm
sgn}\epsilon}, \label{GFpole}
 \ee
where $E_p^2=\epsilon_p^2+\Delta^2$. Integrations over the internal momenta in fermion loops, e.g.,
over $p'$ in~(\ref{gapeq1}), involve energies far off the Fermi surface. One may
renormalize~\cite{M67,FL} the interaction ($\widehat{V}\to \widehat{\Gamma}^\xi$) in such a manner
that integrations go over the region near the Fermi surface and only the quasiparticle (pole) term
in the   Green function (\ref{GFpole}) is operative. Advantage of the Fermi-liquid approach is that
all expressions enter renormalised amplitudes rather than the bare potentials. For
$|\vec{p\,}|\simeq p_{{\rmF}}\simeq |\vec{p\,}'|$ the effective interaction amplitude is a function
of the angle between $\vec{p\,}$ and $\vec{p\,}'$ only. The amplitude in the particle-particle
channel is parameterized as
\be
\big[\widehat{\Gamma}^\xi\big]^{ac}_{bd}=
\Gamma_0^\xi(\vec{n},\vec{n}')(i\sigma_2)^a_b (i\sigma_2)^c_d
+ \Gamma^\xi_1(\vec{n},\vec{n}') (i\sigma_2\,\vec{\sigma})^a_b (\vec{\sigma}\,i\sigma_2)^c_d
,\nonumber\ee
and the interaction in the particle-hole channel is
\be
\big[\widehat{\Gamma}^\om\big]^{ac}_{bd}=
\Gamma_0^\om(\vec{n},\vec{n}')\,\delta^a_b\,
\delta^c_d + \Gamma^\om_1(\vec{n},\vec{n}')\, (\vec{\sigma})^a_b\,
(\vec{\sigma})^c_d .
\nonumber
\ee
Here and below $\vec{n}=\vec{p}/|\vec{p\,}|$ and $\vec{n'}=\vec{p'}/|\vec{p'\,}|$\,. Superscript
"$\om$" indicates that the amplitude is taken for $|\vec{q\,}\vec{v\,}_{{\rm F}}|\ll \om$ and $\om\ll
\epsilon_{{\rm F}}$, where $\om$ and $\vec{q}$ are transferred  energy and  momentum. Amplitudes
$\Gamma_0^{\xi,\om}$, $\Gamma_1^{\xi,\om}$ are expanded in the Legendre polynomials.

Integrating over the internal momenta in loops we can separate the part accumulated in the vicinity
of the Fermi surface $ \int \frac{2\,\rmd^4 p}{(2\pi)^4\,i}\simeq \intop\frac{\rmd \Omega_{\vec
p}}{4\,\pi}\,\times\intop \rmd\Phi_p $ with $\int \rmd\Phi_p = \rho \int_{-\infty}^{+\infty}
\frac{\rmd \epsilon}{2\,\pi\,i} \int_{-\infty}^{+\infty}\rmd \epsilon_p$,
$\rho=\frac{m^*\,p_{{\rmF}}}{\pi^2}$ being the density of states at the Fermi surface. After the
Fermi-liquid renormalization ~(\ref{gapeq1}) reduces to
\be
\Delta(\vec{n}) &=& - A_0\,
\langle \Gamma_0^\xi(\vec{n},\vec{n}') \, \Delta(\vec{n}')
\rangle_{\vec{n}'}\,,
\label{gapeq2}\\
A_0&=&\!\! \int\rmd \Phi_p\, G_{\rm n.s.}(p)\, G^h_{\rm n.s.}(p)\,
\theta(\xi-\epsilon_p)\approx a^2\, \rho\, \ln(2\xi/\Delta),
\nonumber \ee
where we denoted $\langle
\dots\rangle_{\vec{n}}=\int\frac{\rmd \Omega_{\vec n}}{4\pi}\,
(\dots)$ and $\xi\sim \epsilon_{{\rm F}}$\,. One usually
determines the gap supposing $\xi =\epsilon_{{\rm F}}$\,.

\section{Current-current correlator, equations for vertices,  and
vector current conservation}

\subsection{Current-current correlator}

\begin{figure*}
\centerline{\includegraphics[width=0.95\textwidth]{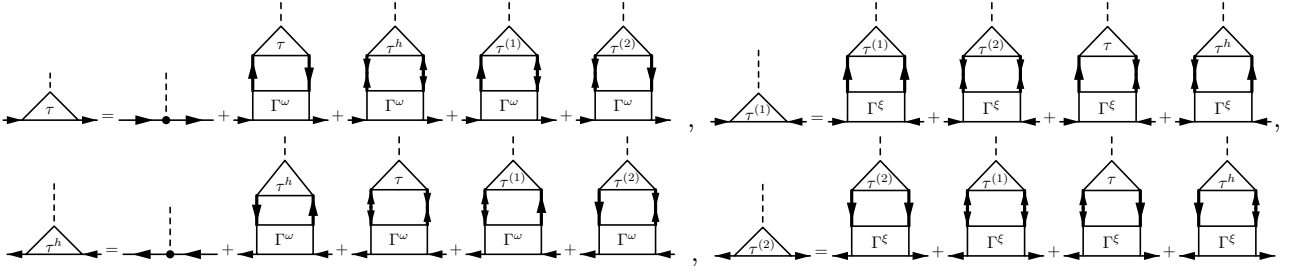}}
\caption{Graphical representation of eqs.~(\ref{LMeq}) \label{fig:verteq}}
\end{figure*}

Applying  theory of Fermi liquids with pairing~\cite{LM63,M67} we can present contributions to the
susceptibility $\chi$ in terms of the diagrams
\be
-i\,\chi=\parbox{6.8cm}{\includegraphics[width=6.8cm]{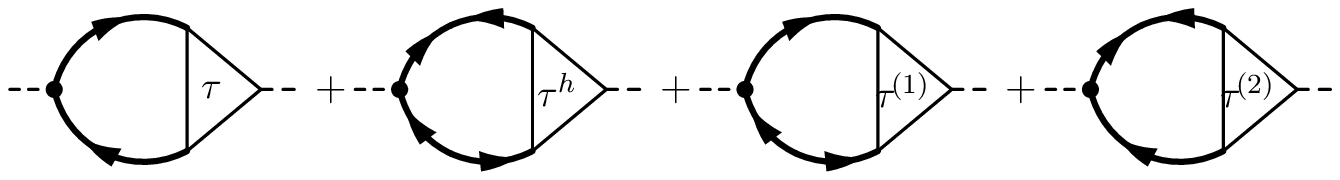}}\,.
\label{chi_diag} \ee
Here dash line relates to the $Z$-boson coupled to the neutral lepton currents, vertices on the
left are the bare vertices following from the Lagrangian (\ref{weak-lag}). The right-hand-side
vertices $\hat{\tau}$, $\hat{\tau}^h$, $\hat{\tau}^{(1)}$ and $\hat{\tau}^{(2)}$ are the full
vertices determined by the diagrams shown in Fig.~\ref{fig:verteq}. The blocks in
Fig.~\ref{fig:verteq} correspond to the two-particle irreducible interaction in the
particle-particle channel, $\Gamma^\xi$, and the particle-hole irreducible interaction  in the
particle-hole channel, $\Gamma^\om$ . We emphasize that only chains of bubble diagrams are summed
up in this particular formulation. Thus, the imaginary part of $\chi$ accounts only for one-nucleon
processes. In order to include two-nucleon processes within a quasi-particle approximation one
should add diagrams with self-energy insertions to the Green functions and iterate the
Landau-Migdal amplitudes $\Gamma^{\om,\xi}$ in Fig.~\ref{fig:verteq} in the horizontal
channel~\cite{VS87,KV95}. In general case for particles with widths
the interpretation of different processes contributing to
$\Im\chi$ is more peculiar and needs another re-summation
scheme~\cite{KV95}.

Taking the imaginary part of $\chi$ we cut a diagram through two fermion lines. Cuts of neutron
lines correspond to neutron PBF processes and those of proton lines correspond to proton PBF
processes. Since the neutron density in a neutron star is much higher than the proton density, we
can drop all diagrams where proton lines are uncut from the set of the bubble chains included in
(\ref{chi_diag}).

Refs.~\cite{FRS76,KHY,YKL99} considered only first two diagrams in~(\ref{chi_diag}) with bare vertices.
Refs.~\cite{VS87,MSTV90} treated those two diagrams with full vertices on the right, whereas
one must consider all four diagrams with the full vertices.

Vector and axial-vector currents contribute to $\chi$ separately,
i.e., $\chi=\chi_V+\chi_A$, where
\be
\chi_a&=& {\rm Tr}\!\! \intop\!\!\frac{d^4 p}{(2\,\pi)^4\, i}\hat{\tau}_a^\om\,\Big\{
\hat{G}_{+}\,\hat{\tau}_a^\dagger\,\hat{G}_{-}
+
\hat{F}^{(1)}_{+}\,\hat{\tau}_a^{h\dagger}\,\hat{F}^{(2)}_{-}
\nonumber\\
&+&
\hat{G}_{+}\,\hat{\tau}_a^{(1)\dagger}\hat{F}^{(2)}_{-}
+
\hat{F}^{(1)}_{+}\,\hat{\tau}_a^{(2)\dagger}\hat{G}_{-} \Big\},\,a=V,A\,.
\label{chi:form}
\ee
Here and below we use the short-hand notations $G_\pm=G(p\pm q/2)$ and the analogous one for the
$F_\pm$ Green functions. All left vertices in (\ref{chi_diag}) are the "bare" vertices,
$\tau^\om_a$, after the Fermi-liquid renormalization~\cite{M67,FL}, $\tau^\om_a=[1+\Gamma^\om_0\,
(G_{+}G_-)^\om]\, \tau_a^0$, which involve the particle-hole effective interaction, $\Gamma^\om_0$,
integrated with off-pole parts of the Green functions $(G_{+}G_-)^\om=\lim_{\vec{q}\to 0}\int
\frac{2\, \rmd^4 p}{(2\,\pi)^4\, i}\,G_{+}G_-$, and $\tau_a^0$ follows from (\ref{weak-lag}). The
difference between $\tau_a^0$ and $\tau_a^\om$ can be cast~\cite{M67} in terms of a local charge of
the quasiparticle $e_{a}=a\, \tau^\om_a/\tau^0_a$\,. Then
\be
\hat{\tau}_V^\om= 
g_V\, \big(\tau_{V,0}^\om\,
l_0-\vec{\tau}_{V,1}^\om\,\vec{l\,}\big),\,
\tau_{V,0}^\om=\frac{e_V}{a}\,,\, \vec{\tau}_{V,1}^\om=\frac{e_V}{a}\,
\vec{v}\,,
\hfill\phantom{xx}
\label{barevert}\\
\hat{\tau}_A^\om= 
-g_A\, \big(\vec{\tau}_{A,1}^\om\vec{\sigma}\,
l_0-\tau_{A,0}^\om\,\vec{\sigma}\vec{l}\,\big),\,
\tau_{A,0}^\om=\frac{e_A}{a}\,,\,
\vec{\tau}_{A,1}^\om=\frac{e_A}{a}\, \vec{v}\,. \nonumber
\ee
For the vector current $e_V=1$ and the vertices $\tau^\om_V$ and $\vec{\tau}^\om_V$ satisfy the
Ward identity $\om\,\tau^\om_{V,0}-\vec{q}\, \vec{\tau}^\om_{V,1}=G_{\rm n.s.}^{\rm
(pole),-1}(p+q/2)-G_{\rm n.s.}^{\rm (pole),-1}(p-q/2)$, with the pole part of the normal state
Green function, $G_{\rm n.s.}^{\rm (pole)}=G_{\rm n.s.}-G_{\rm reg}$. The local charge for the
axial-vector current differs from the unity varying in different parameterizations as $e_A\simeq
0.8\mbox{--} 0.95$, as it follows from studies of the Gamow-Teller transitions in nuclei,
see~\cite{M67,Pyatov83,Borzov03} and references therein.

\subsection{Larkin-Migdal equations for full vertices}

Consider first one sort of nucleons, e.g., neutron. At the Fermi surface the full  vertices
$\hat{\tau}$, $\hat{\tau}^h$, $\hat{\tau}^{(1)}$ and $\hat{\tau}^{(2)}$ can be treated as functions
of out-going momentum $\vec{q}$ and the nucleon Fermi velocity $\vec{v}={v}_{{\rm F}}\vec{n}$,
$\vec{n}=\vec{p}/p$. Their general structures are
\be
\tau_V&=&g_V\, \big(\tau_{V,0}\,
l_0-\vec{\tau}_{V,1}\,\vec{l\,}\big)\,, \quad \tau^h_V=g_V\,
\big(\tau_{V,0}\, l_0+\vec{\tau}_{V,1}\,\vec{l\,}\big)\,,
\nonumber\\
\tau_V^{(1)}&=&-\tau_V^{(2)}=-g_V\,\big(\widetilde{\tau}_{V,0}\,
l_0-\vec{\widetilde{\tau}}_{V,1}\,\vec{l\,}\big)\, i\, \sigma_2\,,
\nonumber\\
\tau_A&=&-g_A\, \big(\vec{\tau}_{A,1}\vec{\sigma}\,
l_0-\tau_{A,0}\,\vec{\sigma}\vec{l}\,\big)\,,
\nonumber\\
\tau^h_A&=&-g_A\, \big(-\vec{\tau}_{A,1}\vec{\sigma}^{\rm T}\, l_0
-\tau_{A,0}\,\vec{\sigma}^{\rm T}\vec{l}\,\big)\,,
\nonumber\\
\tau^{(1)}_A &=& + g_A\, \big(\vec{\widetilde\tau}_{A,1}\,
\vec{\sigma}\, l_0-\widetilde{\tau}_{A,0}\, \vec{\sigma}\,
\vec{l}\, \big)\, i\, \sigma_2\,,
\nonumber\\
\tau^{(2)}_A &=& -
g_A\, i\, \sigma_2\, \big(\vec{\widetilde\tau}_{A,1}\,
\vec{\sigma}\, l_0-\widetilde{\tau}_{A,0}\, \vec{\sigma}\,
\vec{l}\, \big)\,.
\label{fullvert}
\ee
Superscript "$\rm T$" denotes matrix transposition.

As follows from the diagrammatic representation of Fig.~\ref{fig:verteq} the full vertices obey
Larkin-Migdal equations~\cite{LM63}:
\begin{widetext}
\be
\tau_{a,0}(\vec{n},q)=\tau^\om_{a,0}(\vec{n},q)&+&
\big\langle\Gamma_a^{\om}(\vec{n},\vec{n'})\,
\big[ L(\vec{n'},q;P_{a,0})\, \tau_{a,0}(\vec{n'},q)
+
M(\vec{n'},q)\,\widetilde{\tau}_{a,0}(\vec{n'},q)\big]\big\rangle_{\vec{n}'}\,,
\nonumber\\
\widetilde{\tau}_{a,0}(\vec{n},q)=&-&
\big\langle\Gamma_a^{\xi}(\vec{n},\vec{n'})\,
\big[ ( N(\vec{n'},q)+A_0)\, \widetilde{\tau}_{a,0}(\vec{n'},q) + O(\vec{n'},q;P_{a,0})\,
{\tau}_{a,0}(\vec{n'},q)\big]\big\rangle_{\vec{n}'}\,,
\nonumber\\
\vec{\tau}_{a,1}(\vec{n},q)=\vec{\tau}^\om_{a,1}(\vec{n},q)&+&
\big\langle\Gamma_a^{\om}(\vec{n},\vec{n'})\,
\big[ L(\vec{n'},q;P_{a,1})\, \vec{\tau}_{a,1}(\vec{n'},q) +
      M(\vec{n'},q)\,\vec{\widetilde{\tau}}_{a,1}(\vec{n'},q)\big]
\big\rangle_{\vec{n}'}\,,
\nonumber\\
\vec{\widetilde{\tau}}_{a,1}(\vec{n},q)=&-&
\big\langle\Gamma_a^{\xi}(\vec{n},\vec{n'})\,
\big[ ( N(\vec{n'},q)+A_0)\, \vec{\widetilde{\tau}}_{a,1}(\vec{n'},q) + O(\vec{n'},q;P_{a,1})\,
\vec{\tau}_{a,1}(\vec{n'},q)\big]\big\rangle_{\vec{n}'}\,,
\label{LMeq}
\ee
\end{widetext}
where $a=V,A$ and $P_{V,0}=-P_{V,1}=-P_{A,0}=P_{A,1}=1$\,. In
order to write the one set of equations for both vector and axial-vector weak
currents we introduced a new notation for the effective
interaction $\Gamma_a^{\om,\xi}=\Gamma_0^{\om,\xi}$, if $a=V$, and
$\Gamma_a^{\om,\xi}=\Gamma_1^{\om,\xi}$, if $a=A$. Functions $L$,
$M$, $N$, and $O$ are defined as
\be
&&L(\vec{n},q; P)=\int\!\! \rmd \Phi_p\!
\Big[ G_{+} G_--\big(G_{+} G_-\big)^\om - F_{+} F_- P \Big]
\nonumber\\
&&\phantom{L(\vec{n},q; P)}
=a^2\, \rho\,
\Big[\frac{\vec{q}\, \vec{v}}{\om- \vec{q}\,\vec{v}}\,(1- g(z))-g(z)\, (1+P)/2\Big]\,,
\nonumber\\
&&
M(\vec{n},q)= \int\!\! \rmd \Phi_p\!
\Big[ G_{+} F_- - F_{+} G_-\Big]
\nonumber\\
&&\phantom{M(\vec{n},q)}=
-a^2\, \rho\,\frac{\om+\vec{q}\,\vec{v}}{2\, \Delta}\, g(z)\,,
\nonumber\\
&&N(\vec{n},q)= \int\!\! \rmd \Phi_p\!
\Big[G_{+} G^h_-- \big(G_{p} G^h_p\big) \theta(\xi-\epsilon_p)+ F_{+} F_-\Big]
\nonumber\\
&&\phantom{N(\vec{n},q)}=
a^2\, \rho\, \frac{\om^2-(\vec{q}\,\vec{v})^2}{4\, \Delta^2}\, g(z)\,,
\nonumber\\
&&
O(\vec{n},q;P)= -\int\!\! \rmd \Phi_p\!
\Big[ G_{+} F_- + F_{+} G^h_- P\Big]
\nonumber\\
&&\phantom{O(\vec{n},q;P)}=
a^2\, \rho\,
\Big[ \frac{\om+\vec{q}\,\vec{v}}{4\, \Delta}+
\frac{\om-\vec{q}\,\vec{v}}{4\, \Delta}\, P \Big]\, g(z)\,,
\label{depend}\ee
\be
&&\phantom{xxx} g(z^2)=
\intop^{+1/2}_{-1/2}\frac{\rmd x}{4\, z^2\, x^2-z^2+1+i\,0}
\label{gfunc}\\
&&\phantom{xxx g(z^2)}=
 -\frac{{\rm arcsinh}\sqrt{z^2-1}}{z\,
\sqrt{z^2-1}}-\frac{i\, \pi\, \theta(z^2-1)}{2\, z\,
\sqrt{z^2-1}}\,,
\nonumber\\
&&\phantom{xxx}
z^2=\frac{\om^2-(\vec{q}\, \vec{v})^2}{4\, \Delta^2}>1\,,
\quad \vec{v}=v_\rmF \, \vec{n}.
\nonumber \ee

Expressions (\ref{depend}), (\ref{gfunc}) are derived for $T=0$. For finite temperatures,
$T<T_{c}$, all expressions in (\ref{depend}), except for $L$,  hold as well, but with $g(z^2)\to
g(z^2,T)$ and $\Delta \to\Delta (T)$. Generalization of expression for $L$ requires introduction of
one more temperature dependent integral besides $g$. Such expressions were derived by Leggett
in~\cite{Leg65a}. As follows from these expressions  there arises an essential simplification in
the limit of low temperatures, $T\ll \Delta$. We exploit the fact that to calculate the PBF
emissivity we need only imaginary part of the current-current correlator $\Im\chi$. Since $\om
>2\Delta$ for the PBF kinematics, the emissivity is exponentially suppressed by $ e^{-2\Delta/T}$
stemming from the Bose occupation factor $f_{\rm B}(\om)$ in (\ref{emissivity}). Therefore, we may
take $\Im \chi \propto \Im g (T=0)$ since it is already multiplied by the term vanishing for $T\to
0$. Not accounted temperature corrections in $\Im\chi$ prove to be $\sim
1+O(e^{-\Delta/T}(1+T^2/\Delta^2))+O (T^2/\epsilon_{\rm F}^2)$. The latter term follows from the
expansion of Fermi integrals when the integration goes over energy regions far from the Fermi
surface. Such corrections are small in the limit $T\ll \Delta$ and we omit them.

Using vertices (\ref{barevert},\ref{fullvert}) in (\ref{chi:form}) the correlators
$\chi_{V}$ and $\chi_A$ can be expressed as
\be
&&\chi_V(q)\!\!=\!\!g_V^2\, 
\big\langle\big(l_0-\vec{v}\, \vec{l}\,\big)\,
\big(l_0^\dag\, \chi_{V,0}(\vec{n},q) - \vec{\chi}_{V,1}(\vec{n},q)\, \vec{l\,}^\dag\big)
\big\rangle_{\vec{n}},
\nonumber\\
&&\chi_A(q)\!\!=\!\!g_A^2\, 
\big\langle\big(l_0\, \vec{v}-\vec{l}\,\big)\, \big(l_0^\dag\,
\vec{\chi}_{A,1}(\vec{n},q) - \chi_{A,0}(\vec{n},q)
\vec{l\,}^\dag\big) \big\rangle_{\vec{n}},
\nonumber\\
&&\chi_{a,0}(\vec{n},q)\!\!=\!\! L(\vec{n},q;P_{a,0})\,
\tau_{a,0}(\vec{n},q)+M(\vec{n},q)\,
\widetilde{\tau}_{a,0}(\vec{n},q),
\nonumber\\
&&\vec{\chi}_{a,1}(\vec{n},q)\!\!=\!\! L(\vec{n},q;P_{a,1})\,
\vec{\tau}_{a,1}(\vec{n},q)+M(\vec{n},q)\,
\vec{\widetilde{\tau}}_{a,1}(\vec{n},q).
\nonumber 
\ee

\subsection{Solution  for vector and axial-vector parts of the current-current correlator}

It is natural to expect that first and higher Legendre harmonics
of $\Gamma^{\om,\xi}_{0,1}(\vec{n},\vec{n'})$ are smaller than the
zero-th ones due to the centrifugal factor~\cite{M67}. This allows
us to retain only zero harmonics
$\Gamma^{\om,\xi}_{0,1}(\vec{n},\vec{n'})=\Gamma^{\om,\xi}_{0,1}={\rm const}$\,,
 expressed through dimension-less Landau-Migdal parameters as
~\cite{M67} $\Gamma_0^{\om,\xi}= f^{\om,\xi}/(a^2\rho(n_0))$ and
$\Gamma_1^{\om,\xi}= g^{\om,\xi}/(a^2\rho(n_0))$. The values of
parameters are extracted from the analysis of atomic nucleus
experiments~\cite{M67,Pyatov83,Borzov03} or in some approximations can be
calculated starting from a microscopic nucleon-nucleon
interaction~\cite{HSF07}.
Actually, in isospin asymmetric matter $f^{\om}$ and
$g^{\om}$ are different for interactions between two neutrons ($f_{nn}^{\om}$, $g^{\om}_{nn}$), two
protons ($f_{pp}^{\om}$, $g^{\om}_{pp}$) and neutron  and proton ($f_{np}^{\om}$, $g^{\om}_{np}$).
Note that values $f_{nn}^{\om}$,  $f_{pp}^{\om}$ are necessarily positive, the requirement of the
stability of the nucleon matter, whereas corresponding values in the particle-particle channel
$f_{nn}^{\xi}$,  $f_{pp}^{\xi}$ are negative, otherwise there would be no $1S_0$ pairing. In this
respect our derivations differ from   those  which do not distinguish interactions in particle-hole
and particle-particle channels and use Nambu-Gorkov formulations with one bare potential ($V<0$ in
our case).

For the angular-independent amplitudes
(only zero-th harmonics are included)
 the Larkin-Migdal equations
(\ref{LMeq}) get simple solutions:
\be
&&\tau_{a,0}(q) =
\gamma_a(q;P_{a,0})\,\tau^\om_{a,0}\,,
\nonumber\\
&&\gamma_a^{-1}(q;P)=1 -
\Gamma^\om_a\,\langle\mathcal{L}(\vec{n},q;P)\rangle_{\vec{n}}\,,
\nonumber\\
&&\mathcal{L}(\vec{n},q;P)= L(\vec{n},q;P)-
\frac{\langle  O(\vec{n},q;P)  \rangle_{\vec{n}}}{\langle
N(\vec{n},q)\rangle_{\vec{n}}}\, M(\vec{n},q) ,
\nonumber\\
&&\widetilde{\tau}_{a,0}(q) = -\frac{\langle
O(\vec{n},q;P_{a,0})\rangle_{\vec{n}}}{\langle
N(\vec{n},q)\rangle_{\vec{n}}}\, \tau_{a,0}(q)\,.
\label{solLMeq}
\ee
We have exploited here the relation
$1=-\Gamma^\xi_0\, \langle A_0 \rangle$ following from the gap
equation (\ref{gapeq2}). Although  integrals in (\ref{LMeq}) do
not produce terms $\propto \vec{v}$ for constant
$\Gamma^{\om,\xi}_a$, the vector vertices $\vec{\tau}_{a,1}$ and
$\vec{\widetilde{\tau}}_{a,1}$ gain new terms proportional to
$\vec{q}$, thus,
$\vec{\tau}_{a,1}(\vec{n},q)=\vec{\tau\,}^\om_{a,1}(\vec{n},q)+
\vec{n}_q\, \tau_{a,1}^{(q)}(q)$ and
$\vec{\widetilde{\tau}}_{a,1}(\vec{n},q)= \vec{n}_q\,
\widetilde{\tau}_{a,1}^{(q)}(q)$, where
$\vec{n}_q=\vec{q}/|\vec{q}\,|$ and
\be
&\tau^{(q)}_{a,1}(q)=\gamma_a(q;P_{a,1})\, \Gamma_a^\om\,
\langle \widetilde{\mathcal{L}}(\vec{n},q;P_{a,1})\, (\vec{n}\,\vec{n}_q) \rangle_{\vec{n}}\,,&
\label{tauq}\\
&\widetilde{\tau}_{a,1}^{(q)}(q)= -\frac{\langle O(\vec{n}, q;P_{a,1})\rangle}
{\langle N(\vec{n}, q)\rangle_{\vec{n}}}\, {\tau}_{a,1}^{(q)}
-\frac{\langle O(\vec{n}, q; P_{a,1}) (\vec{n}\,\vec{n}_q)
\rangle_{\vec{n}}}{\langle N(\vec{n}, q)\rangle_{\vec{n}}},
&
\nonumber\\
&\widetilde{\mathcal{L}}(\vec{n},q;P) =
L(\vec{n},q;P)- \frac{\langle M(\vec{n},q) \rangle_{\vec{n}}}
{\langle N(\vec{n},q)\rangle_{\vec{n}}}\, O(\vec{n},q;P) \,. \nonumber &
\ee
With (\ref{barevert},\ref{solLMeq},\ref{tauq}) we cast
$\chi_a^\mu=(\chi_{a,0},\vec{\chi}_{a,1})$ as
\be
&&\chi_{a,0}(\vec{n},q) = \gamma_a(q;P_{a,0})\,
\mathcal{L}(\vec{n},q;P_{a,0})\,, \label{chi01}\\
&&\vec{\chi}_{a,1}(\vec{n},q) =
\vec{v}\,\gamma_a(q;P_{a,1})\,{\mathcal{L}}(\vec{n},q;P_{a,1})
+\delta\vec{\chi}_{a,1}(\vec{n},q)\,, \nonumber\\
&&\delta\vec{\chi}_{a,1}(\vec{n},q) = \frac{M(\vec{n},q)}{\langle
N(\vec{n'}, q)\rangle_{\vec{n'}}}\, \langle O(\vec{n'}, q;
P_{a,1}) (\vec{v}-\vec{v\,}') \rangle_{\vec{n'}} \nonumber\\ && +
\mathcal{L}(\vec{n},q;P_{a,1}) \gamma_a(q;P_{a,1})\,\Gamma_a^\om\,
\langle\widetilde{\mathcal{L}}(\vec{n'}, q; P_{a,1})
(\vec{v}\,'-\vec{v}\,) \rangle_{\vec{n'}}\,. \nonumber \ee Here
$\gamma_a$ are precisely those nucleon-nucleon correlation factors
that have been introduced in \cite{VS87}. They depend on
Landau-Migdal parameters in the particle-hole reaction channels.

\subsection{Vector current conservation}

Now we are in the position  to verify that the  correlator of the vector current $\chi_V^\mu$
supports the current conservation. First we note that there are convenient relations
\be
&&\langle \mathcal{L}(\vec{n},q;P)-
\widetilde{\mathcal{L}}(\vec{n},q;P)\rangle_{\vec{n}}= 0,
\nonumber
\\ && \langle \om\, \mathcal{L}(\vec{n},q;\pm 1) -
\widetilde{\mathcal{L}}(\vec{n},q;\mp 1)
\,(\vec{q}\,\vec{v}\,)\rangle_{\vec{n}} = 0\,,
\nonumber 
\\
&&\langle\vec{q}\,\vec{\chi}_{a,1}(\vec{n},q)\rangle_{\vec{n}}
=\gamma_a(P_{a,1})\langle
(\vec{q}\,\vec{v}\,)\,\widetilde{\mathcal{L}}(\vec{n},q;P_{a,1})\rangle_{\vec{n}},
\nonumber
\\
&&\big\langle (\vec{q}\,\vec{v}\,)\,
\big(\om\,\mathcal{L}(\vec{n},q;+1)-
(\vec{q}\,\vec{v}\,)\,\mathcal{L}(\vec{n},q;-1) \nonumber\\
&&-\vec{q}\,\delta\vec{\chi}_{V,1}(\vec{n},q,\Gamma^\om=0) \big)
\big\rangle_{\vec{n}}=\vec{q}^{\,2}{a^2\, n}/{m^*}\,.
\label{urel:4}
\ee
These relations help us to establish important properties of the
vector current correlators (\ref{chi01}):
\be
\langle \om\,
\chi_{V,0}-\vec{q}\,\vec{\chi}_{V,1}\rangle_{\vec{n}}=
\gamma_V (q;+1)\, \gamma_V (q;-1)\, \om\,\Gamma_V^\om
\phantom{\,}
\nonumber\\
\times
\langle \mathcal{L}(\vec{n},q;+1) \rangle \big( \langle
{\mathcal{L}}(\vec{n},q;-1) \rangle-\langle
{\mathcal{L}}(\vec{n},q;+1)\rangle\big)
\nonumber\\
=O(f^\om\,g\,\vec{q\,}^6\, v_\rmF^6/\om^6)\,,
\phantom{xxxxi\,}
\nonumber\\
\Im\langle
(\vec{q}\,\vec{v}\,)\, (\om\,
\chi_{V,0}-\vec{q}\,\vec{\chi}_{V,1})\rangle_{\vec{n}}
=
\om\,\Gamma_V^\om\,\langle \mathcal{L}(\vec{n},q;+1) \rangle
\phantom{xxxxx\,}
\nonumber\\
\times
\langle (\vec{q}\,\vec{v}\,)\big[\gamma_V(q;+1) \mathcal{L}(\vec{n},q;+1)
-  \gamma_V (q;-1) \mathcal{L}(\vec{n},q;-1)\big] \rangle
\nonumber\\ =O(f^\om\,g\,\vec{q\,}^6\,
v_\rmF^6/\om^6)\,.\phantom{xxxxx} \label{Ward}
 \ee
We use here expansion of $\mathcal{L}$ and $\widetilde\mathcal{L}$
in the series for $|\vec{q}|\, v_\rmF /\om\ll 1$:
\be
\mathcal{L}(\vec{n},q;+1) &=& \frac{y\,x}{1-y\,x}+ g\,
y^2({\textstyle \frac13}-x^2) \nonumber\\ &\times& \big(1+y\, x +
y^2\,({\textstyle\frac13} + x^2) \big)+O(g\,y^5 )\,,
\nonumber\\
\mathcal{L}(\vec{n},q;-1) &=& \frac{y\,x}{1-y\,x}-
g\,y\,x\, \big(1+y\,x+y^2\, x^2+y^3\, x^3\big)
\nonumber\\
&+& O(g\,y^5)\,,
\nonumber\\
\widetilde{\mathcal{L}}(\vec{n},q;+1)  &=& \frac{y\,x}{1-y\,x} -g\, y\,x\, (1+y^2\,x^2)
\nonumber\\
&-&g\, y^2\, ({\textstyle \frac13}-x^2)\, \big(1-y^2\,({\textstyle \frac13}+x^2)\big) +O(g\,y^5)\,,
\nonumber\\
\widetilde{\mathcal{L}}(\vec{n},q;-1) &=& \frac{y\,x}{1-y\,x} -g\, y^2\, x^2\, (1+ y^2\, x^2 )
\nonumber\\
&- &g\, x\, y^3\, ({\textstyle \frac13}-x^2) +O(g\,y^6)\,,
\label{expand}\\
y&=&q\, v_\rmF/\om\,, x=(\vec{q}\,\vec{v}\,)/|\vec{q}|\,
v_\rmF\,,\quad
\nonumber\\
g&=&g\big(\frac{\om^2}{\Delta^2\,} (1-y^2\,x^2)\big)\,.
\nonumber
\ee
Relations (\ref{Ward}) demonstrate that the imaginary part of the
vector current correlator calculated with  full vertices
(\ref{solLMeq},\ref{tauq}) is transverse,
\be
\Im \langle\tau^\om_\mu\,\chi_{V}^\nu\rangle_{\vec{n}}q_\nu
=O(f^\om\,g\,\vec{q\,}^6\, v_\rmF^6/\om^6 ),
\label{trans} \ee
at least up to terms of the
higher order than $f^\om (|\vec{q}|\,v_\rmF/\om)^5 g$\,,
which are beyond the Fermi-liquid approximation for the Green
functions (\ref{GF},\ref{GFpole}). This ensures conservation of
the vector current. Note that for $\Gamma^\om =0$ or  $g=0$ we
have $\Im \langle\tau^\om_\mu\,\chi_{V}^\nu\rangle_{\vec{n}}q_\nu \equiv 0$
and then the vector current is conserved exactly. Since  $g\propto
\Delta (T)$ and $\Delta = 0$ for $T>T_c$, we have proven in
passing that the vector current is conserved exactly above $T_c$.

In order to prove the transversality of the real part of the vector-current correlator it would be
necessary to include the tadpole diagram contribution
\parbox{1cm}{\includegraphics[width=1cm]{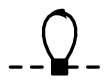}}, where the coupling originates from the "gauging" of the
kinetic term, $\psi \vec{\nabla\,}^2\psi/2\, m^*$, of an effective
non-relativistic nucleon Hamiltonian.

Some comments on the approximations done in previous works would be here appropriate. In all
previous works the vector current contribution has been considered as the dominant term for the
case of $1S_0$ pairing. Expressions for the PBF emissivity in~\cite{FRS76,KHY,YKL99} can be
recovered, if we put $\Im\chi_{V,0}(\vec{n},q)=\Im L(\vec{n},q;+1)$ and $\Im\vec{\chi_{V,1}}=0$\,.
Result of \cite{VS87,MSTV90} is obtained by taking $\Im\chi_{V,0}(\vec{n},q)=\Im
L(\vec{n},q;+1)/(1+\Gamma^\om_0\,\Re L(\vec{n},q;+1))^2$ and also $\Im\vec{\chi_{V,1}}=0$\,.
Setting $\Gamma^\om =0$ in the limit $\om\ll \Delta$ and $\vec{q}=0$ we reproduce expressions of
Ref. \cite{LP}.

Note that relations (\ref{urel:4}) do not hold with
$\mathcal{L}$ replaced by $L$ and, hence, the transversality relation (\ref{trans}) is spoiled, if
one ignors the anomalous vertex terms.

\section{Neutrino emission via neutron PBF}

After correlators (\ref{chi01}) are established it remains to
take the sum over the lepton spins and integrate over the leptonic
phase space in (\ref{emissivity}). The latter can be easily done
with the help of the Lenard integral~\cite{Lenard}
\be
\intop\frac{d^3 q_1}{(2\pi)^3\,2\, \om_1}
\frac{d^3 q_2}{(2\pi)^3\,2\, \om_2} \,\sum\{l^\mu\,
l^{\dag\nu}\}\,\delta^{(4)}(q_1+q_2-q)
\nonumber\\
=\frac{1}{48\, \pi^5}\,\big(q^\mu\, q^\nu-g^{\mu\nu}\,
q^2\big)\, \theta(\om)\, \theta(\om^2-\vec{q\,}^2)\,.
\label{Lenard}
\ee
Now the neutrino emissivity (\ref{emissivity}) can be cast as
\be
\varepsilon_{\nu\nu} &=&
\varepsilon_{\nu\nu,V}+\varepsilon_{\nu\nu,A}\,,
\nonumber\\
\varepsilon_{\nu\nu,a} &=& \frac{G^2}{8}\,
g_a^{*\,2}\int^\infty_0\rmd \om\, \om\, f_{\rm B}(\om)
\int_0^\om\frac{\rmd |\vec{q}|\, \vec{q\,}^2}{6\,\pi^4}\,\frac{\kappa_a}{a^2}
\nonumber\\
&=&\frac{G^2\,g_a^{*\,2}}{240\, \pi^4}\int^\infty_0\rmd \om\,
\om^6\, f_{\rm B}(\om) \,Q_a(\om),
\label{emiss}\\
Q_a(\om) &=& \frac{5}{\om^5} \intop_0^\om\rmd |\vec{q}|\vec{q\,}^2
\frac{\kappa_a}{a^2}\,,
\label{defQa}\\
\kappa_a &=& \intop \frac{d^3 q_1}{2\, \om_1} \frac{d^3 q_2}{2\, \om_2}
\,\delta^{(4)}(q_1+q_2-q)
\nonumber\\
&\times&\frac{3}{4\,\pi}\, \Im\sum\ \chi_a(q).
\label{defKa}\ee
In $\kappa_a$ the sum is taken over the lepton spins. Shortening notations we introduced in
(\ref{emiss}) effective couplings $g_a^{*}=e_a\,g_a$. The particular normalization of the quantity
$Q_a$ is chosen so that for $Q_a(\om)=Q^{0}(\om)$ with
\be
Q^{(0)}(\om)&=&-\rho\, \Im g(\om^2/4\Delta^2)\, \label{Q0} \ee we
obtain  expression for the neutron PBF emissivity
\be
\epsilon_{\nu\nu}^{(0n)}= \frac{4 \rho_n\,G^2 \Delta^7_n}{15\,
\pi^3}I(\frac{\Delta_n}{T}),\, I(z)=\!\intop^\infty_1\! \frac{\rmd
y\, y^5}{\sqrt{y^2-1}}e^{-2zy}\,,
\label{emiss0} \ee
which coincides with the old result ~\cite{FRS76,VS87} after the replacement $e^{-2zy}
\to\frac{1}{(e^{zy}+1)^2}$. From now on we restore, where it is necessary, subscripts "$n$" or
"$p$" to distinguish neutron and proton PBF processes, respectively.

\subsection{Emissivity on vector current}

For the vector current we have
\be
\kappa_V&=&\Im\Big[\vec{q\,}^2\,
\langle\chi_{V,0}(\vec{n},q)\rangle_{\vec{n}}+
\langle(\vec{q}\,\vec{v}\, )\, \vec{q}\,
\vec{\chi}_{V,1}(\vec{n},q)\rangle_{\vec{n}} \nonumber\\ &+&
(\om^2-\vec{q\,}^2)\, \langle\vec{v}\,
\vec{\chi}_{V,1}(\vec{n},q)\rangle_{\vec{n}} \nonumber\\ &-& \om\,
\langle(\vec{q}\,\vec{v}\,)\,
\chi_{V,0}(\vec{n},q)\rangle_{\vec{n}}- \om\, \langle\vec{q}\,
\vec{\chi}_{V,1}(\vec{n},q)\rangle_{\vec{n}}\Big].
\label{kappaV:full} \ee

Using relations (\ref{chi01}) and (\ref{Ward}) we can simplify
(\ref{kappaV:full}) as
\be
\kappa_V&=&(\vec{q\,}^2-\om^2)\,
\Im \langle\chi_{V,0}(\vec{n},q) -
 \vec{v}\, \vec{\chi}_{V,1}(\vec{n},q)\rangle_{\vec{n}}.
\label{kappaV:Ward}
\ee
Both scalar and vector components in
(\ref{kappaV:Ward}) are of the order $v_\rmF^4$,
\be
\Im\langle\chi_{V,0}(\vec{n},q) \rangle &\approx& -\frac{4\,
\vec{q\,}^4 v_\rmF^4}{45\, \om^4}\,a^2\, \rho\, \Im
g(\frac{\om^2}{4\Delta^2})>0\,, \nonumber\\ \Im\langle
\vec{v}\,\vec{\chi}_{V,1}(\vec{n},q) \rangle\, &\approx&
-\frac{2\,\vec{q\,}^2 v_\rmF^4}{9\om^2}\,a^2\, \rho\, \Im
g(\frac{\om^2}{4\Delta^2})>0\,. \nonumber
 \ee
We have put $\gamma_V\to 1$ since $\gamma_V \simeq
1+O(f_{nn}^{\om} v_{\rm F}^2)$.

Note that the first term in (\ref{kappaV:Ward}), $\propto \Im\chi_{V,0}$, would give a negative
contribution to $Q_V$. Only due to the presence of  {\em the vector component of the vertex}, second term
in (\ref{kappaV:Ward}), the full expression for the reaction probability becomes positive. This is
because we used Ward identities, which impose relations between zero- and vector components.
However, if one keeps in (\ref{kappaV:full}) only the first term related to the zero-th component
of the vertex and drops other terms, as it was done in early works, the expression for the reaction
probability would be also positive.

Then in terms of $Q_V$ we get
\be
Q_V^n \simeq \frac{4}{81}\, v_{\rmF, n}^4\, Q^{(0n)}(\om)\,.
\label{QV} \ee Finally for the neutron PBF emissivity on the
vector current we obtain (for one neutrino flavor) \be\label{eV}
\epsilon_{\nu\nu,V}^{\rm nPFB}&\simeq&\epsilon_{\nu\nu}^{(0n)}\,
g_V^2\,\frac{4}{81}v_{\rmF,n}^4 \,, \ee

Note that in spite of Ref. \cite{LP} used approximation  $\om \ll
\Delta_n$, which  is not fulfilled in PBF case, our expression
(\ref{eV}) only slightly deviates from the corresponding result
obtained in \cite{LP}.

Authors of Ref.~\cite{SMS} calculated the susceptibility $\chi_V$ including only the zero-th
component, $\chi_{V,0}$, for $v_\rmF=0$, performing an expansion for small $\vec{q\,}$\,. They
found the leading term $\propto \vec{q\,}^2/2\,m$.
However, it has the opposite sign (see (48) in \cite{SMS})
to the second term, $\propto I_B$, in (40), that would yield reaction probability in case, if bare
vertices were used. Note also that the key equations (35,38-45) in \cite{SMS} differ from
Larkin-Migdal equations (\ref{depend}) (for $T\ll \Delta$ as supposed in \cite{LM63}, and for
$v_{\rm F}=0$ as assumed in \cite{SMS}). As it follows from (\ref{depend}) and (\ref{solLMeq}) our
expression for $\langle \mathcal{L}(\vec{n},q,+1)\rangle_{\vec{n}}\propto q^4\, v_{\rmF,n}^2$
vanishes, if $v_{\rmF,n}\to 0$\,, although $\vec{q}^{\,2}/m^*$ terms were present in the original
loop integrals.

\subsection{Emissivity on axial-vector current}

Now we focus on  the process going on the axial-vector current. Then
\be
\kappa_A&=&\Im\Big[ \vec{q\,}^2 \langle \vec{v}\,
\vec{\chi}_{A,1}(\vec{n},q)\rangle_{\vec{n}}+(3\,\om^2-2\,
\vec{q\,}^2)\, \langle \chi_{A,0}(\vec{n},q)\rangle_{\vec{n}}
\nonumber\\ &-&\om \langle \vec{q}\,
\vec{\chi}_{A,1}(\vec{n},q)\rangle_{\vec{n}} -\om\,
\langle(\vec{q}\,\vec{v}\,)\,
\chi_{A,0}(\vec{n},q)\rangle_{\vec{n}} \Big]\,. \label{kappaA}
 \ee
The last two  crossing terms in the squared brackets cannot be
eliminated. Keeping only terms $\propto v_\rmF^2$
we cast (\ref{kappaA}) as
\be
\kappa_A &=&\Im\Big[ \vec{q\,}^2\, v_\rmF^2 \, \langle
L(\vec{n},q;+1)\rangle_{\vec{n}} + (3\, \om^2-2\,\vec{q\,}^2) \langle
L(\vec{n},q;-1)\rangle_{\vec{n}} \nonumber\\ &-&\om^2 \langle
L(\vec{n},q;-1)\rangle_{\vec{n}} -\om\, \langle (\vec{q}\,
\vec{v\,})\, L(\vec{n},q;-1)\rangle_{\vec{n}} \Big]
\nonumber\\
&\approx& - a^2\,\rho\,v_\rmF^2\, \vec{q\,}^2\,
\big[1+(1-{\textstyle\frac23\, \frac{\vec{q}^{\,2}}{\om^2}})-
{\textstyle \frac23}\big]\, \Im g(\frac{\om^2}{4\Delta^2}).
\label{kappaA:2}
\ee
As in case with the vector current, simplifying we could put $\gamma_A =1$ since $\gamma_A \simeq
1+O(g_{nn}^{\om} v_{\rm F}^2)$.

The contribution of the axial-vector current to the neutrino emissivity is determined by
\be
Q_A^n(\om)\simeq \big(1+\frac{11}{21}-\frac23\big)\,v_{\rmF,n}^2\,Q^{(0n)}(\om).
\label{QA}
\ee
Second  term  in round brackets of ~(\ref{QA}) has been mentioned already in~\cite{FRS76} and then
recovered in~\cite{YKL99}. Our coefficient ($11/21$) is twice larger than that presented in those
works. We notice that the integral $I_s /2 =(u' v-v' u)^2$, where $u$ and $v$ are coefficients of
the  Bogolyubov transformation, is in~\cite{YKL99} twice as large as that in~\cite{FRS76}. In
agreement with the former evaluation, we arrive at the coefficient $11/21$ rather than at $11/42$,
as presented in~\cite{FRS76,YKL99}. The first term in (\ref{QA}) (for $m^* =m$) is the same, as in
Ref.~\cite{YKL99}, which calculated this relativistic correction for the fist time. The factor
$(m^* /m)^2$ does not arise in our calculations, since the mass renormalization is performed
everywhere, including the vertices. Otherwise the Ward identity  would not hold for the
renormalized "bare" vertex $\tau^\om_\mu$\,. The third term related to the time-space component
product was not considered before.

Finally for the neutron PBF emissivity on the axial-vector current
we obtain (for one neutrino flavor)
\be\label{eA}
\epsilon_{\nu\nu,A}^{\rm nPBF}&\simeq& \frac67\, g_A^{*\,2}\,
v_{\rmF,n}^2\,\epsilon_{\nu\nu}^{(0n)}\,.
\ee
The resulting
emissivity is the sum of contributions (\ref{eV}) and (\ref{eA}),
\be
\label{eAtotn}
\epsilon_{\nu\nu}^{{\rm nPBF}}= \epsilon_{\nu\nu,V}^{\rm pPBF}
+\epsilon_{\nu\nu,A}^{{\rm nPBF}}\simeq \epsilon_{\nu\nu,A}^{{\rm nPBF}}.
\ee
The
axial-vector term, being $\propto v_{\rmF}^2$, is now the
dominating contribution. Thus, the ratio of the emissivity of the
neutron PFB obtained here to the emissivity calculated
in~\cite{FRS76,YKL99}, where main contribution was due to the
vector current, is
\be
R({\rm nPFB})=\frac{\epsilon_{\nu\nu}^{{\rm nPBF} }}{\epsilon_{\nu\nu}^{(0n)}}\simeq
\frac{6}{7} \, g_A^{*\,2}\, v_{\rmF,n}^2 = F_n\,v_{\rmF,n}^2\,.\nonumber
 \ee
For $n=n_0=0.17~{\rm fm}^{-3}$, $m^* =0.8~m$, we estimate $F_n
\simeq 0.9\mbox{--} 1.2$, $v_{\rmF,n}\simeq 0.36$ and $R\simeq
0.12\mbox{---}0.15$. For $n=2n_0$, $m^* =0.7m$, $R$ increases up
to $0.24\mbox{---} 0.32$. This is in drastic contrast with
estimations $R(\rm nPFB)\sim 10^{-3}$ in~\cite{LP}  (being
actually valid only for the rate of partial vector current
contributions, rather than for the full emissivities) and $R(\rm
nPFB)\simeq 5\cdot 10^{-3}$ obtained in~\cite{SMS}.

\section{Neutrino emission via proton PBF}

Now we turn to the proton PBF processes. If protons were the only particles in the neutron star
medium, we could use the results obtained above for the neutron PBF and just replace $g_V^{(n)}
\rightarrow g_V^{(p)}= c_V$, $v_{\rmF,n}\to v_{\rmF,p}$, $f_{nn}\to f_{pp}$ and $g_{nn}\to g_{pp}$,
$e_a^n\to e_a^p$.

\subsection{Emissivity on vector current}

Since the bare vertex yields now $\big(g_V^{(p)}\big)^2=c_V^2
\simeq 0.002$ compared to $\big(g_V^{(n)}\big)^2 =1$ in the
neutron case one could naively think that the emissivity of the
proton PBF process on the vector current is suppressed by a factor
$\sim 10^{-3}(\Delta_p /\Delta_n)^{13/2}e^{2(\Delta_p
-\Delta_n)/T}$ compared to the emissivity of the neutron PBF
process. For imaginary purely proton matter we would find in the vector channel
\be\label{eVp}
\epsilon_{\nu\nu,V}^{\rm }&\simeq&\epsilon_{\nu\nu}^{(0p)}\, c_V^2
\,\frac{4}{81}v_{\rmF, p}^4 \,,
\\
\epsilon_{\nu\nu}^{(0p)}&=& \frac{4 \rho_p\,G^2 \Delta^7_p}{15\,
\pi^3}I(\frac{\Delta_p}{T})\,.
 \ee

In addition to that the emissivity is already suppressed in the vector channel by factor
$\frac{1}{20}v_{\rmF,p}^4$, Leinson and Perez~\cite{LP} found  extra suppression. They
included interaction of protons via photons. This produces new contributions to the susceptibility
$\chi$ of the following type
\be
\includegraphics[width=5cm]{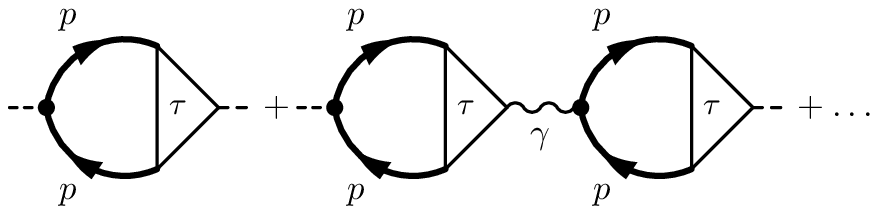}
\nonumber \ee
Dotts assume  infinite summation of the bubble
chains with all four types of the vertices shown in
(\ref{chi_diag}). The wavy line is the dressed photon Green
function. Simplifying, Ref. \cite{LP} used the static Coulomb
potential instead. To illucidate the origin of differences in
our estimations with those in~\cite{LP} we will use the same approximations.
Effectively the summation leads to the replacement~\cite{LP}
\be\label{newv}
\tau_{V,0}\to \frac{\tau_{V,0}}{\epsilon_{\rm C}(q)},
\ee
where $\epsilon_{\rm C} (q) \simeq 1 + {\om_{\rm pl}^2}/{\om^2}$ is the dielectric constant,
$\om_{\rm pl}^2 =4\, \pi\,e^2 n_p /m^*$ is the proton plasma frequency with $e^2=1/137$. Setting
$m^*=m$, $\om \simeq 2\Delta_p$, ($\om\simeq 2\Delta +O(T)$ for the PFB processes), $\Delta\simeq
1.76 T_c$ and $T_c\sim 1$~MeV for $x_p =n_p/n \sim 0.03$ at $n=n_0$, cf. Fig. 2 of \cite{Army}, we
obtain $\epsilon_{\rm C} (q) \simeq 1+ 0.3\sim 1$ in disagreement with estimation of \cite{LP},
where applying their result to the neutron star matter authors put $n_p =n_0$ and $\om \simeq T_c$
that resulted in the estimation $\epsilon_{\rm C} (q)\sim 10^{2}$. Thus, a suppression factor $<
10^{-6}$ of the emissivity of the proton PBF process quoted in \cite{LP} is misleading.  Note that
correction of the vertex (\ref{newv}) affects only the process on the vector current, since photon
is the vector particle.

For the neutron star matter, the replacement (\ref{newv})
is not sufficient, since protons are embedded
into the electron liquid of the same concentration and into a much
more dense neutron liquid. Renormalization of the weak vector
interaction of protons in this case can be taken into account, if
we replace the bare coupling $\tau^\om_V$ as follows
\be
\parbox{8cm}{\includegraphics[width=8cm]{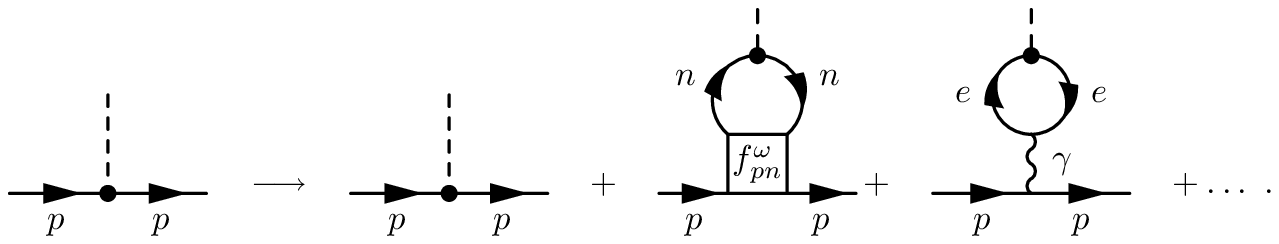}}
\label{ppvertV}
\ee
Dots stand for other graphs not shown
explicitly, like $\Delta (1232)n$-loop term, etc. Simplifying, we
ignore these rather small correction terms. The second graph has
been incorporated in \cite{VS87,MSTV90} that results in the shift
\be
c_V \to c_V -f^\om_{np}\rho^{-1}(n_0)\Re L_{nn}\,\gamma(f^\om_{nn}),
\nonumber
\ee
where
$\gamma^{-1}(f^\om_{nn})=1-f^\om_{nn}\rho^{-1}(n_0)\Re L_{nn}$ and $L_{nn}=\langle
L(\vec{n},q;g=0) \rangle_{\vec{n}}/a^2 =\rho\langle
\vec{q\,}\vec{v} /(\om-\vec{q\,}\vec{v})\rangle_{\vec{n}}$ is the
Lindhard function. This correction (although $\propto v_{\rm
F}^2$) leads to the strong enhancement of the tiny bare vertex. A
numerically larger correction comes from the
electron--electron-hole polarization term (the third graph). Such
a possibility has been discussed in~\cite{VKK} for the process of
a possible massive photon decay and then it was taken into account
in the proton PBF emissivity in~\cite{L00}. Altogether these
corrections can be incorporated to the resulting expression for
the emissivity of the proton PBF process with the help of the
replacement, cf.~\cite{V01},
\be c_V^2 \to {\cal{F}}_p
\simeq\epsilon_{\rm C}^{-2} (q) \Big[f^\om_{np}\rho^{-1}(n_0) \Re
L_{nn}\gamma(f^\om_{nn})+0.8\, C_{ve}\Big]^2.
\ee
Here $C_{ve}=1$ is
the electron weak vector coupling. Thus we find
\be\label{eVpp}
\epsilon_{\nu\nu,V}^{\rm pPBF}&\simeq
&\epsilon_{\nu\nu}^{(0p)}\,{\cal{F}}_p \,\frac{4}{81}v_{\rmF, p}^4
\,, \ee
where the pre-factor ${\cal{F}}_p\sim 1$. Finally we
obtain an estimate
\be
R_V [p/n]=\frac{\epsilon_{\nu\nu,V}^{\rm pPBF}}{\epsilon_{\nu\nu,V}^{\rm nPBF}}\sim
x_p^{4/3}\Big(\frac{\Delta_p}{\Delta_n}\Big)^{13/2}\,e^{2\,(\Delta_n
-\Delta_p)/T}.
\ee
The ratio $R_V[p/n]$ is sensitive to the values of the
proton and neutron gaps as functions of density, $\Delta_{n,p}(n)$,
the proton fraction $x_p$, and the temperature $T$.

\subsection{Emissivity on axial-vector current}

Now we consider axial-vector channel. Photon exchange does not
contribute in this channel. The main correction to the vertex
comes from the iteration of the $nn$-loops:
\be
\parbox{8cm}{\includegraphics[width=8cm]{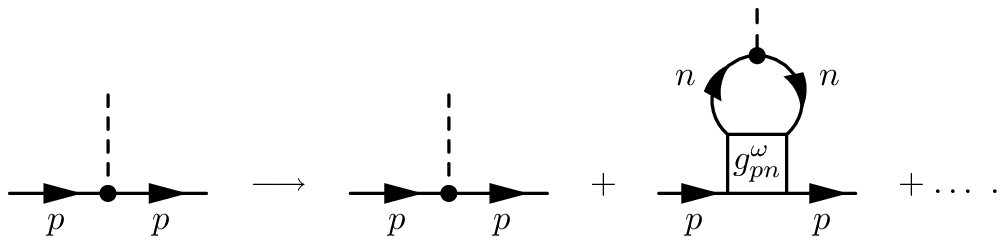}} \label{ppvertV}
\ee
Simplifying, we will suppress correlation
factors, such as $\gamma^2 (g^\om_{nn})\simeq 1+O(g_{nn}^{\om}v_{{\rm
F},n}^2)$. Thus we obtain
\be\label{eApp} \epsilon_{\nu\nu,A}^{\rm
pPBF}&\simeq&\epsilon_{\nu\nu}^{(0p)}\,\, \frac67 \,
g_A^{*\,2}\, v_{\rmF,p}^2\,.
\ee
Comparison with \cite{YKL99} can be done quite
similar to that performed above for neutrons.

We conclude \be\label{eAtot} \epsilon_{\nu\nu}^{\rm
pPBF}=\epsilon_{\nu\nu,V}^{\rm pPBF}+\epsilon_{\nu\nu,A}^{\rm
pPBF}\simeq \epsilon_{\nu\nu,A}^{\rm pPBF}.\ee

For the ratio $R[p/n]$ we find
\be
 R[p/n]&=&
\frac{\epsilon_{\nu\nu}^{\rm pPBF}}{\epsilon_{\nu\nu}^{\rm
nPBF}}\simeq \frac{\epsilon_{\nu\nu,A}^{\rm
pPBF}}{\epsilon_{\nu\nu,A}^{\rm nPBF}} \nonumber\\ &\sim&
x_p^{4/3}\Big(\frac{\Delta_p}{\Delta_n}\Big)^{13/2}\,e^{(\Delta_n
-\Delta_p)/T}.
 \ee
The ratio $R[p/n]$ is sensitive to the choice of pairing  gaps,
temperature and the proton fraction $x_p$ and can be both $\lsim 1$ and $\gsim 1$\,.

\section{Conclusions}

In this paper we re-calculated neutrino emissivity via neutron and proton pair breaking and
formation processes. We used Larkin-Migdal-Leggett Fermi-liquid approach to strongly interacting
systems with the pairing. Compared to the Nambu-Gorkov formalism, the Larkin-Migdal-Leggett
approach allows for different interactions in the particle-particle and the particle-hole channels,
as it is the case for nuclear matter.

To be specific we focused  our discussion on the $1S_0$ pairing. We support statement of~\cite{LP}
that medium effects essentially modify vector current vertices. Only the careful account of these
effects allows to fulfill the Ward identity and to protect conservation of the vector current.
Compared to the emissivity calculated in~\cite{FRS76} the partial contribution to the emissivity on
the neutron vector current proved to be dramatically suppressed, roughly by a factor $\sim
0.1\times v_{\rmF,n}^4$, $v_{\rmF,n}$ is the neutron Fermi velocity, cf.~\cite{LP}. A similar
suppression factor arises for the partial contribution to the emissivity on the proton vector
current, provided one replaces neutron Fermi velocity by the proton Fermi velocity.
Electron--electron-hole and neutron--neutron-hole polarization effects play a crucial role in the
latter estimation. Proton--proton-hole polarization effects are suppressed (these statements are at
variance with the estimations in~\cite{LP}).

The dominating contribution to the neutron and proton pair breaking and formation emissivity comes
from the weak axial-vector current. Finally, the neutron pair breaking and formation emissivity
proves to be suppressed compared to that of~\cite{FRS76} by a factor $\sim 0.12\mbox{--} 0.15$ at
nuclear saturation density and $\sim 0.24\mbox{--} 0.32$ at twice  nuclear saturation density. For
the proton pair breaking and formation the emissivity deviates only by a factor close to unity from
the expression used previously in \cite{YKL99}. These our findings differ from those
in~\cite{LP,SMS}, where authors concluded that the neutron and proton pair breaking and formation
emissivities are dramatically suppressed.  Modifications of the neutron and proton pair breaking
and formation reaction rates that we have found are
 probably not strong enough to essentially influence on  the values of
surface temperatures of neutron stars computed previously.

One may rise the question how much emissivity of other relevant neutrino processes might be
changed, if  medium effects in presence of nucleon pairing are correctly included. Although we did
not perform corresponding cumbersome calculations let formulate our conjectures:

In the reactions with charged currents, as the direct Urca and the modified Urca processes, the
transferred neutrino energy is $\om \simeq \mu_e =p_{\rmF,p}\gg 2\Delta$. Therefore the anomalous
Green functions are taken in the limit $\om \gg 2\Delta$. In this limit $g$-function tends to zero
(as it follows from the corresponding asymptotic in (\ref{gfunc})). Effects of normal correlations
and pion softening were evaluated in~\cite{VS86,VS87,MSTV90,V01}, resulting in significant
enhancement of two-nucleon reaction rates. Specifics of the superfluid matter manifest themselves
in reactions with charged currents  mainly through the phase-space suppression factors.

The two-nucleon bremstrahlung processes going on the neutral currents are similar to the pair
breaking and formation reactions. In a normal phase the emissivities are governed by the
axial-vector current~\cite{MUrca}. Thereby, we do not expect a strong suppression of these rates in
a superfluid phase except the natural phase-space suppression estimated in  previous works. As in
the case  of the two-nucleon reactions going on charged currents, nucleon short-range correlations
and pion softening significantly influence the reaction rates, cf. ~\cite{VS86,MSTV90}.

Our findings are relevant also for  calculations of the quark-pair breaking and formation processes
and other quark propagation processes in the color-superconducting medium, which use bare-loop
results, e.g. see Refs.~\cite{JP,Sh}.  Note that since the pairing gaps in the color
superconductors can be rather large, $2\Delta>\mu_e$, both reaction rates  on neutral and charged
currents (Urca) might be affected. Ref.~\cite{KR} considered neutrino scattering off breaking pairs
in color-flavor-locked medium within the Nambu-Gorkov formalism. However, they included only the
zero-th component of the vertex, and their expressions for the current-current correlator
in the non-relativistic limit do not coincide with the Larkin-Migdal-Leggett expressions  that we
have reproduced above.

An interesting observation was made recently in Ref.~\cite{GBSMK}. A natural explanation of the
super-burst ignition would require a strong suppression of the neutron pair breaking and formation
emissivity for low baryon densities. For $n\sim 10^{12}\,\mbox{g}/\mbox{cm}^3$ we
estimate a suppression factor as $\sim 0.1\times (n/n_0)^{2/3}\sim 0.003$.

Another relevant issue is related to absorption and scattering processes of  low-energy ($\om
\lsim$ few MeV) neutrinos and antineutrinos on nuclei. In absence of the electron Fermi sea the
correlation effects may manifest themselves in reactions on both neutral and charged currents.
There are different sources for neutrinos of such energies, e.g., reactor neutrinos are good
candidates. The Sun and supernova neutrinos also have pronounced low energy tails. The observation
of supernova neutrinos might provide us with unique information on the core-collapse and on the
compact star formation and cooling \cite{Volpe}. Geo-neutrinos and antineutrinos from the progenies
of $\mbox{U}$, $\mbox{Th}$ and $^{40}\mbox{K}$ decays inside the Earth bring to the surface
information from the whole planet, about its content of radioactive elements \cite{Bellini}.
Finally, verifying the existence of the relic neutrino sea with temperature $
T_{\nu}/T_{\gamma}=(4/11)^{1/3}$ represents one of the main challenges of the modern cosmology
\cite{Lazauskas}.

From a general point of view, our results strongly support the conclusion of Refs.
\cite{VS86,MSTV90,KV99,V01} about the essential role played by different medium effects in the
neutrino evolution of neutron stars, as it was demonstrated in the framework of the "nuclear-medium
cooling scenario"~\cite{SVSWW97,BGV04,GV05}. Without a proper inclusion of medium effects it is
difficult to reach sound conclusions. Further investigations in this direction are required.

\acknowledgements

We are grateful to D. Blaschke, I.N.~Borzov, L.V.~Grigorenko,
Yu.B.~Ivanov, B.~Friman, D. Page, S.V.~Tolokonnikov and
D.G.~Yakovlev for the discussions. This work was supported in part
by the US Department of Energy under contract No.
DE-FG02-87ER40328, by the Deutsche Forschungsgemeinschaft  DFG,
project 436 RUS 113/558/0-3, and by the Russian Foundation for
Basic Research, grant RFBR 06-02-04001.

\appendix
\section{Calculation with bare vertices in the loop}

Let us now comment on results of previous calculations of the
neutron PBF emissivity. If, as in calculations~\cite{FRS76,YKL99},
one neglects contributions from the anomalous vertices
$\widetilde\tau$, i.e. $M\to 0$ and $\mathcal{L}\to L$ in
~(\ref{chi01}), and puts $\gamma_a=1$, one may recover results of
calculations with bare vertices ("b.v.").  Expression (\ref{kappaV:full}) for
$\kappa_V$ becomes
\be
\kappa_V^{\rm b.v.}&=&\Im\Big[ \vec{q\,}^2
\,\langle{L}(\vec{n},q;+1)\rangle_{\vec{n}} \nonumber\\
&+&v_\rmF^2\,
\langle\big(\om^2+(\vec{q}\,\vec{n})^2-\vec{q\,}^2\big)\,
L(\vec{n},q;-1)\rangle_{\vec{n}} \nonumber\\ &-&\om\,\langle
(\vec{q}\,\vec{v}\,)\big({L}(\vec{n},q;+1)+L(\vec{n},q;-1)\big)\rangle_{\vec{n}}\Big]\,.
\label{kappaV} \ee
In the axial-vector channel the corresponding quantity is
$\kappa_A^{\rm b.v.}\propto v_{\rmF}^2$.
Therefore we will keep also the $v_\rmF^2$ corrections in (\ref{kappaV}).
In  calculations~\cite{FRS76,YKL99} such terms were dropped.
The second term in (\ref{kappaV}) can be indeed
neglected, as being $\propto v_\rmF^4$, since
$\langle{L}(\vec{n},q;-1)\rangle_{\vec{n}}\propto v_{\rm F}^2$.
The third crossing term of the order $v_{\rm F}^2$
disregarded in previous calculations can be dropped only,
if conditions (\ref{Ward}) hold, that is not the case for the bare vector
current-current correlator. We keep this term here. For
$|\vec{q\,}|\, v_\rmF\ll \om$ (i.e. for $v_\rmF\ll 1$ since
$|\vec{q\,}|\lsim \om$) we get
\be
\frac{\kappa_V^{\rm b.v.}}{a^2\, \rho} &\approx& -\vec{q\,}^2\,\Im
g\Big(\frac{\om^2-(\vec{v\,}\vec{q\,})^2}{4\, \Delta^2}\Big)
-\frac{\vec{q\,}^4\, v_\rmF^2}{3\ \om^2}\,\Im
g\Big(\frac{\om^2}{4\,\Delta^2}\Big)
\nonumber\\
&+& \frac23\,\vec{q\,}^2\, v_\rmF^2\,
\Im g\Big(\frac{\om^2}{4\,\Delta^2}\Big).
\label{kappaV:exp} \ee
The first line in (\ref{kappaV:exp}) comes from the expansion of the first term in (\ref{kappaV}).
The first term in the second line follows from  the crossing term in (\ref{kappaV}). Using
(\ref{kappaV:exp}) we calculate
\be
Q_V^{\rm b.v.} &=&\Big(1+\frac{5}{21}\, v_\rmF^2-\frac23\,
v_\rmF^2\Big)\, Q^{(0)}(\om)\,.
\nonumber \ee
Dropping the $v_\rmF^2$ corrections and using expression (\ref{Q0})
for $Q_V^{(0)}$ one may reproduce the
vector current contribution to the emissivity obtained in
Refs.~\cite{FRS76,YKL99} (and \cite{VS87,MSTV90}, provided one
sets there $\gamma_V =0$).

The expression for the emissivity of neutron PBF on the
axial-vector current calculated with the bare vertices in the loop
does not deviate from that in (\ref{QA}) derived above with the
full vertices. Note, however, that second and third terms in (\ref{QA}) differ from those
used in previous calculations .



\end{document}